\documentclass[%
preprint,
showpacs,
showkeys,
nofootinbib,
amsmath,amssymb,amsfonts,
aps,
prd,
floatfix,a4paper,11pt]{revtex4-1}
\usepackage{}
\usepackage{color}
\usepackage{CJK}
\usepackage{graphicx}% Include figure files
\usepackage{dcolumn}% Align table columns on decimal point
\usepackage{bm}% bold math
\usepackage[dvipdfm,bookmarks=true,colorlinks,%
            citecolor=blue,linkcolor=blue, hypertex, %
            breaklinks=true]{hyperref}

\begin{document}
\title{Structure function of holographic quark-gluon plasma: Sakai-Sugimoto model versus its non-critical version}
\date{\today}
\author{Yan Yan Bu}
\email{yybu@itp.ac.cn}
\author{Jin Min Yang}
\email{jmyang@itp.ac.cn}
\affiliation{Institute of Theoretical Physics, Academia Sinica, Beijing 100190, P. R. China}

\begin{abstract}
Motivated by recent studies of deep inelastic scattering (DIS) off the $\mathcal{N}=4$ super-Yang-Mills (SYM) plasma, holographically dual to $AdS_5\times S^5$ black hole, we in this note use the spacelike flavor current to probe the internal structure of one holographic quark-gluon plasma, which is described by the Sakai-Sugimoto model at high temperature phase (i.e., the chiral symmetric phase). The plasma structure function is extracted from the retarded flavor current-current correlator. Our main aim here is to explore the effect of non-conformality on these physical quantities. As usual, our study is under the supergravity approximation and the limit of large color number. Although the Sakai-Sugimoto model is non-conformal, which makes the calculations more involved than the well-studied $\mathcal {N}$=4 SYM case, the result seems to indicate that the non-conformality has little essential effect on the physical picture of the internal structure of holographic plasma, which is consistent with the intuition from the asymptotic freedom of QCD at high energy. While the physical picture underlying our investigation is same as the DIS off the $\mathcal {N}$=4 SYM plasma with(out) flavor, the plasma structure functions are quantitatively different, especially their scaling dependence on the temperature, which can be recognized as model-dependent. As a comparison, we also do the same analysis for the non-critical version of the Sakai-Sugimoto model which is conformal in the sense that it has a constant dilaton vacuum. The result for this non-critical model is much similar to the conformal $\mathcal{N}=4$ SYM plasma. We therefore attribute the above difference to the effect of non-conformality of the Sakai-Sugimoto model.

\end{abstract}
\pacs{11.25.Tq, 12.38.Lg, 12.38.Mh}
\keywords{Gauge/Gravity Duality, Deep Inelastic Scattering, Holographic Dual of QCD}

\maketitle

\section{Introduction} \label{section1}
In heavy ion collision, which is now experimentally studied at RHIC and LHC, the so-called quark-gluon plasma seems to be strongly interacting and behaves like perfect liquid \cite{hep-ph/0312227}, which is greatly different from the previously recognized weakly-coupled one. This brings non-perturbative investigations of hadronic matter at high temperature and high density produced in heavy ion collision into an urgent stage. Though the lattice method can calculate some properties of strongly-coupled system, it is still constrained to extract some static quantities such as hadron mass spectrum and
thermodynamical behavior. What is worse is that when adding finite density or chemical potential into the thermal QCD, the lattice calculation usually confronts the notorious sign problem. Therefore, improvement in theoretical understanding of strongly-coupled quark-gluon plasma (sQGP) should not only go beyond the traditional perturbative QCD (pQCD) approach but also reveal some properties out of equilibrium, such as transport
properties, dispersion relation, high-energy scattering and so on.

Gauge/gravity duality \cite{hep-th/9711200,*hep-th/9802109,*hep-th/9802150} states that
strong coupling gauge theory can be mapped to weak coupling gravity with a negative cosmological constant in the limit of large 't Hooft coupling and large $N_c$, where $N_c$ is the number of color of gauge theory. Although the gravity dual of the realistic QCD has not been established till now, one expects this duality to be of great importance in understanding some non-perturbative properties of QCD or at least some universal features of strongly-coupled system. In fact, over the last decade, using gauge/gravity duality technique to study properties of sQGP has gained great success
\cite{0804.2423,*0811.3113,*0812.0500,*0901.0935,*1003.3291,*1101.0618}. However, most of the studies have focused on static or hydrodynamic properties at large-scale or long-time compared to the inversion of the temperature of the system. Therefore, it is of great interest to study hard probe of the plasma and reveal its internal structure, which should be in contrast with the parton picture of a single hadron in pQCD. Studies on this topic\footnote{For studies on deep inelastic scattering under the technique of
gauge/gravity duality, see, e.g., refs \cite{hep-th/0207071,*hep-th/0207144,*hep-th/0501038,*hep-th/0603115, *hep-ph/0610168,
*0710.2148,*0711.0221,*0712.3530,*0804.1562,*0805.4346,*0807.1917,*0904.2870,*0907.4604,
*0912.4333,*0912.4704} and \cite{PhysRevLett.105.072003,*1001.3548,*1002.0566,
*1003.3066,*1007.0306,*1007.2448,*1007.4362,*1010.2815,*1105.2907,*1105.2999}.}
was originally proposed in \cite{hep-th/0109174,*hep-th/0209221} for $\mathcal{R}$-current scattering off a dilaton hadron in hard-wall model and later generalized to the plasma case without flavor in \cite{0710.5297} and with flavor in \cite{0912.0231,*0912.2238}.

The main lesson\footnote{For details on discussions of the structure function and the partonic picture of holographic quark-gluon plasma, see \cite{0710.5297,0912.0231,*0912.2238,0803.2481}.} from these investigations is that at high energy the $\mathcal{R}$-current or flavor current probes the partonic behavior of the plasma, giving nonvanishing plasma structure function; while at low energy the current is not absorbed by the plasma, indicating vanishing contribution to the plasma structure function\footnote{When taking into account the non-perturbative tunneling effect of current when encountering a narrow and high potential barrier, an exponentially suppressed structure function can be obtained.}. Besides the plasma structure function, another important quantity is the so-called plasma saturation momentum, which is the critical energy defining the transition from weakly quasi-elastic scattering to high-energy deep inelastic scattering. In other words, the partonic picture for holographic plasma exists only when all partons have transverse momenta below the saturation momentum.

Since previous studies are mainly focusing on the D3-brane geometry, which is dual to the conformal $\mathcal{N}$=4 SYM, we in this note\footnote{As in the literature, we here for simplification do not consider charge density which can be modeled by time component of flavor gauge field in present construction and just leave this problem for further investigation.} use one non-conformal gravity dual model of QCD to probe the effect of the non-conformality on the plasma structure. To be specific, the model under consideration here is the transversely-intersecting D4/D8/$\overline{\text{D8}}$ brane
system, which is now usually referred to as the Sakai-Sugimoto model \cite{PTP-113-843,*hep-th/0507073}. It is one of the most successful holographic QCD models from the top-down approach in realizing some phenomena of low energy QCD such as confinement/deconfinement phase transition\footnote{Recently, there appears one paper \cite{1107.4048} which proves that the previous interpretation of transition between the $AdS$ soliton and the black D4 brane as the strong coupling continuation of the confinement/deconfinement transition in four dimensional Yang-Mills theory is not valid. The authors there proposed an alternative gravity dual of the confinement/deconfinement transition. For details on this topic, see the original work \cite{1107.4048}.}, non-Abelian chiral symmetry breaking and vector meson dominance. The temperature under gauge/gravity duality approach can be realized by extending the color brane geometry to a black hole \cite{hep-th/9803131}. The high temperature phase means that the chiral symmetry is restoring, which is denoted by parallel profile of flavor D8/$\overline{\text{D8}}$ branes. Note that the temperature will be much smaller than the four-momentum of the flavor current, since in this note we are focusing on high energy scattering to probe the internal structure of holographic plasma. The setup in this note is similar to \cite{0912.0231,*0912.2238}, but due to the fact that the induced metric on the flavor worldvolume reduces to $AdS_5 \times S^4$ black hole, it is essential the same as that of the $\mathcal{R}$-current DIS off the $\mathcal{N}$=4 SYM plasma. Thus, the procedure here is similar to \cite{0710.5297} and much simpler than \cite{0912.0231,*0912.2238}. To extract the structure function, we need to study the flavor current propagation in holographic plasma. According to gauge/gravity duality, this can be achieved by studying the U(1) flavor gauge field in curved spacetime. The corresponding action and equation of motion determining this dynamics are encoded in a Maxwell theory in curved five dimensional spacetime. On the other hand, the parton structure function has standard definition in quantum field theory, which is encoded in the retarded current-current correlator. One main task of gauge/gravity duality is to calculate the retarded Green function for finite temperature field theory from dual gravity following the recipe in \cite{0205051}. In this sense, the information of the structure function is totally encoded in the solution of the above-mentioned Maxwell equation in five dimensional curved spacetime of asymptotic AdS type. One additional key point under this prescription is that the solution should obey incoming wave boundary condition at the horizon, reflecting the full-absorption characteristic of black hole. Once the equation of motion for the flavor U(1) gauge field is turned into Schr$\ddot{\text{o}}$dinger type, it will be found that their behavior is very similar to that of the $\mathcal{N}$=4 SYM case. This is natural since the setup of holographic models is very general. This may also be recognized as one universal characteristic of gauge/gravity duality.

Another motivation for our study in this note is to probe some universal features of the structure of holographic plasma, described by the intersecting D-brane system in conformal or non-conformal case. We have learned that the procedure and the physical picture for these calculations are universal, which is in part due to the unified approach of gauge/gravity duality in dealing with strongly-coupled problems. The specific form of the structure function is model-dependent, which may allow us to resort to the experiments to judge which model is much closer to physical reality. For the two structure functions, we find that in the non-conformal background they also satisfy the Callan-Gross relation in the limit of large Bjorken variable being defined later, which is same as in the $\mathcal{N}$=4 SYM case. Therefore, it is reasonable to conjecture that this relation should be universal for holographic plasma described by the intersecting D-brane system, having nothing to do with the conformality of holographic background.

Since the flavor gauge field we are focusing on in this note can also be considered as the gravity dual of vector meson, our study can then be regarded as the completion of previous studies of mesonic quasinormal modes in \cite{PhysRevD.77.126008}. Here, we extend these studies to high momentum and high frequency limit, in contrast to previous hydrodynamic behavior or just high frequency limit. However, here we will not go into the detailed numeric extraction of mesonic quasinormal frequency in high frequency and high momentum limit, and leave this task for future direction.

We will also do the same analysis for the non-critical version of the Sakai-Sugimoto model for comparison. This non-critical model was proposed in \cite{hep-th/0403254,*hep-th/0411009,*hep-th/0510110} to overcome some drawbacks of the critical Sakai-Sugimoto one, which are general in critical string theory. This model can be thought of as a conformal one, so we expect the related results in this model should be more similar to the $\mathcal{N}$=4 SYM case. This expectation has been confirmed in the fact that the structure function for non-critical model has the same scaling dependence on temperature as that of $\mathcal{N}$=4 SYM case. This should be another difference between the Sakai-Sugimoto model and its non-critical version.

Another striking point is that the stringy imprints have appeared in the final results of the structure functions for both models considered in this note. Seemingly, this would sway the successful aspects of the Sakai-Sugimoto model. However, if we recall that the holographic plasma is quite different from the realistic sQGP, we should just take these unsatisfactory features as non-universal ones and focus on some universal features emergent from different holographic models.

The rest of this note is organized as follows. In Sec.~\ref{section2} we first give a brief overview on the Sakai-Sugimoto model as well as its non-critical version, and then give the basic equations for later calculations. In Sec.~\ref{section3} we state detailed extraction of the structure function for the Sakai-Sugimoto model and list the main results for its non-critical version. Then we have a brief discussion about the results. Sec.~\ref{section4} is devoted to a short summary and some open questions.

\section{Overview of models and basic equations}
\label{section2}
In this section we first recapitulate the Sakai-Sugimoto model and its non-critical version \cite{PTP-113-843,*hep-th/0507073,hep-th/0403254,*hep-th/0411009,
*hep-th/0510110,0711.4273}. Then we turn to the definitions for the plasma structure function in terms of physical quantity in standard field theory. We will also state basic equations determining the flavor current propagation in these plasmas from the viewpoint of dual gravity, which are essential for later extraction of the structure function. We will follow the notation conventions in \cite{0710.5297,0912.0231,*0912.2238}.

\subsection{Sakai-Sugimoto model versus its non-critical version}
The bulk background geometry of the Sakai-Sugimoto model is given by a ten-dimensional supergravity description of $N_c$ coincident D4-branes in type-IIA superstring theory compactified on a circle. According to \cite{hep-th/9803131}, there are two different metrics for this supergravity, representing two different phases of holographic QCD. The transition between these two phases is interpreted as the deconfinement phase transition. Here, we just focus on the high temperature deconfining phase, described by following backgrounds\footnote{Due to the periodic identification of time coordinate, it should be the Euclidean time. However, we here use the Minkowskian signature for later convenience of extraction of Minkowski space retarded Green function.}
\begin{eqnarray}
&& ds^2=\left(\frac{u}{R}\right)^{\frac{3}{2}}\left(-f(u)dt^2+d \vec{x}^2+dx_4^2\right)+
\left(\frac{R}{u}\right)^{\frac{3}{2}}\left(\frac{du^2}{f(u)}+u^2d\Omega^2_4\right),\\
&& e^{\phi}=g_s \left(\frac{u}{R}\right)^{\frac{3}{4}}, \quad
f(u)=1-\left(\frac{u_T}{u} \right)^3, \quad
t\sim t+\beta=t+\frac{4\pi R^{\frac{3}{2}}}{3u_T^{\frac{1}{2}}}.
\end{eqnarray}
The temperature of holographic plasma dual to the above background is
\begin{equation}
T=\frac{1}{\beta}=\frac{3u_T^{\frac{1}{2}}}{4\pi R^{\frac{3}{2}}}.
\end{equation}
The curvature radius $R$ of the background is related to the string coupling $g_s$ and
string length $l_s$ via
\begin{equation}
R^3=\pi g_s N_c l_s^3\equiv \pi\lambda l_s^3.
\end{equation}
Here in the second equality we have defined the 't Hooft coupling constant $\lambda$ from the dual gravity side as $\lambda\equiv g_s N_c$.

The above gravity background is dual to the gluon sector, and the quark sector can be introduced in quenched approximation by adding $N_f$ pairs of D8 and $\overline{\text{D8}}$ flavor branes to the above geometry and make them transverse to the circle along $x_4$. In the quenched limit, $N_f\ll N_c$, the backreaction of the flavor branes on the background geometry can be neglected. Dynamics of the flavor sector is encoded in the Dirac-Born-Infeld (DBI) plus the Chern-Simons (CS) actions for the flavor branes in above background. However, the CS-term will be exactly zero in this note as there is no background for the U(1) gauge field on the flavor branes.

Chiral phase transition for the flavor sector in this deconfined phase can happen and it has a beautiful geometric explanation: parallel profile of the flavor D8 and $\overline{\text{D8}}$ branes stands for chiral restoring phase while connected U-shaped profile for chiral broken phase. In general, high temperature corresponds to chiral restoring phase while low temperature for chiral broken phase (the details on this topic can be found in \cite{hep-th/0604161,*hep-th/0604173}). In the high temperature phase, the induced metric on the flavor branes has the following standard $AdS$ form
\begin{equation}
ds_8^2=\left(\frac{u}{R}\right)^{\frac{3}{2}}\left(-f(u)dt^2+d \vec{x}^2\right)+
\left(\frac{R}{u}\right)^{\frac{3}{2}}\left(\frac{du^2}{f(u)}+u^2d\Omega^2_4\right).
\end{equation}

The other model we are concerning here is the non-critical version of the above one.
It is based on the supergravity description of $N_c$ coincident D4-branes in six dimensions with one dimension compactified on a circle as in the Sakai-Sugimoto model. The corresponding background geometry takes the following form in high temperature phase
\begin{eqnarray}
&& ds^2=\left(\frac{u}{R}\right)^2\left(-f(u)dt^2+d\vec{x}^2+dx_4^2 \right)
  + \left(\frac{R}{u}\right)^2\frac{du^2}{f(u)},\\
&& e^\phi=\frac{2\sqrt{2}}{\sqrt{3}N_c},\quad R^2=\frac{15}{2},\quad
f(u)=1-\left(\frac{u_T}{u}\right)^5 ,\quad t\sim t+\beta=t +\frac{4\pi R^2}{5u_T}.
\end{eqnarray}
Flavor quark sector can be introduced by adding $N_f$ pairs of D4 and $\overline{\text{D4}}$ branes into above background geometry. One important feature of this model is that it does not have the undesired internal space, which may introduce the unwanted KK modes. Another striking characteristic of this model is that it has constant dilaton vacuum as for the D3-brane geometry in ten dimensions, which should signal that this model is a conformal one. As in the Sakai-Sugimoto model, chiral restoring phase means induced metric on the flavor branes takes the form
\begin{equation}
ds_4^2=\left(\frac{u}{R}\right)^2\left(-f(u)dt^2+d\vec{x}^2\right)+\left(\frac{R}{u}
\right)^2\frac{du^2}{f(u)}.
\end{equation}

For convenience in later calculations, we now rescale the radial coordinate $u$ to make it dimensionless by a transformation $u_T/u=r$. Then the induced geometry on the flavor branes takes the following simplified versions:
\begin{eqnarray}
&& ds^2_8=\left(\frac{u_T}{R}\right)^{\frac{3}{2}}r^{-\frac{3}{2}}\left(-f(r)dt^2+
d\vec{x}^2\right)+R^{\frac{3}{2}}u_T^{\frac{1}{2}}r^{-\frac{5}{2}}
\left(\frac{dr^2}{f(r)}+r^2d\Omega^2_4\right),\label{induced metric for ss} \\
&& e^{\phi}=g_s \left(\frac{u_T}{R}\right)^{\frac{3}{4}}r^{-\frac{3}{4}},\qquad f(r)=1-r^3;\\
&& ds_4^2=\left(\frac{u_T}{R}\right)^2\frac{1}{r^2}\left(-f(r)dt^2+ d\vec{x}^2\right)
  +\left (\frac{R}{r}\right)^2\frac{dr^2}{f(r)},\\
&& e^\phi=\frac{2\sqrt{2}}{\sqrt{3}N_c},\quad f(r)=1-r^5.
\end{eqnarray}
Note that under the new coordinates, the interval for radial coordinate $r$ is located in a finite regime:~ $r\in[0,1]$, and the horizon is now at $r=1$ while boundary at $r=0$, which will make later analysis convenient.

Before concluding this subsection we give a short comment about one general feature of above models. Whether there is background on flavor branes, fluctuations of the flavor gauge field and scalar mode in chiral restoring phase will always decouple, which together with exact $AdS$ forms of the induced metrics will greatly simplify later calculations. This is also one reason why we choose the transversely intersecting D-brane systems for study. As mentioned in Sec.~\ref{section1}, the plasma structure function is completely encoded in the dynamics of the flavor U(1) gauge field propagating through the above-mentioned geometry, which is described by the DBI actions on the flavor branes.

\subsection{DIS: field theoretical definitions}
\label{sub DIS}
Deep inelastic scattering in QCD is a powerful tool in exploring the hadron structure. Here we mainly focus on the electromagnetic mediation between the charged lepton and the hadron. The basic objective of DIS is to compute the retarded current-current correlator defined by
\begin{equation}
\Pi_{\mu\nu}(k)\equiv i \int d^4x e^{-ik\cdot x}\theta(x_0)\langle\left[J_\mu(x),J_\nu(0)\right]\rangle,\label{polarization1}
\end{equation}
where $k$ is the four-momentum of the electromagnetic current, $\langle...\rangle$ means quantum vacuum expectation, and $J_{\mu}(x)$ is the mediated electromagnetic current. When considering the lepton scattering off the plasma, the hadron should be replaced by the plasma system and the vacuum polarization tensor Eq.~(\ref{polarization1}) is modified to
\begin{equation}
\Pi_{\mu\nu}\left(k,T\right)\equiv i \int d^4x e^{-ik\cdot x}\theta(x_0)\langle\left[J_\mu(x),J_\nu(0)\right]\rangle_T,
\label{polarization2}
\end{equation}
where $\langle...\rangle_T$ means thermal expectation value in the plasma system.

Though the gravity dual of the $SU(N)\times U(1)_{e.m.}$ gauge theory has not been established, we can use a non-dynamical electromagnetic field to model the photon just as in condensed matter physics given that electromagnetic coupling constant is very small. In the present context, the electromagnetic current is replaced by the flavor U(1) current and also denoted as $J_{\mu}(x)$. Now we list the general structure of the thermal polarization tensor defined in Eq.~(\ref{polarization2}). According to gauge symmetry and rotation symmetry of thermal field theory, it can be decomposed into two scalar functions as
\begin{eqnarray}
\Pi_{\mu\nu}\left(k,T\right)=\left(\eta_{\mu\nu}-\frac{k_{\mu}k_{\nu}}{Q^2}\right)
\Pi_1\left(x,Q^2\right)+\left(n_{\mu}-k_{\mu}\frac{n\cdot k}{Q^2}\right)
\left(n_{\nu}-k_{\nu}\frac{n\cdot k}{Q^2}\right)\Pi_2\left(x,Q^2\right),
\label{tensor decom}
\end{eqnarray}
where $\eta_{\mu\nu}=\left(-1,1,1,1\right)$, $Q^2\equiv k^{\mu}k_{\mu}$ is the current virtuality, and $n_{\mu}$ is the four-velocity of the plasma which will be chosen as $n_{\mu}=\left(-1,0,0,0\right)$ to signal that the plasma is at rest. We have also defined the Bjoken variable as $x=-Q^2/[2\left(n\cdot k\right)T]$. Then the DIS structure function of the plasma can be extracted from the polarization tensor as
\begin{equation}
F_1\equiv\frac{1}{2\pi}\text{Im} \Pi_1, \quad F_2\equiv\frac{-\left(n\cdot k\right)}{2\pi T}\text{Im}\Pi_2.
\label{structure}
\end{equation}

For sake of later convenience, we now proceed to express the above two functions in terms of longitudinal and transverse polarization tensors $\Pi_{LL}$ and $\Pi_{yy}$ being introduced in Sec.~\ref{section3}:
\begin{eqnarray}
F_1&=&\frac{1}{2\pi}\text{Im}\Pi_{yy}=\frac{1}{2\pi}\text{Im}\Pi_{zz},\label{F1}\\
F_2&=&\frac{\omega^2}{q^2}\left(\frac{Q^2 x}{\pi}\text{Im}\Pi_{LL}+2x F_1\right).\label{F2}
\end{eqnarray}
In obtaining these two expressions we have used the flavor current momentum defined in Eq.~(\ref{eq:A}) and assumed that the plasma is at rest. From Eq.~(\ref{F2}) we can see that in the interesting kinematic regime, $\omega^2/q^2=1-Q^2/q^2\simeq1$, $F_2$ can be simplified further to
\begin{equation}
F_2\simeq\frac{Q^2 x}{\pi}\text{Im}\Pi_{LL}+2x F_1. \label{F2'}
\end{equation}
If the first term in Eq.~(\ref{F2'}) could be negligible, we then
straightforwardly come to the familiar Callan-Gross relation
$F_2\simeq 2x F_1$.

In particle physics, the structure function has been studied by the operator product expansion technique for specific hard processes. The parton model is suitable for weak coupling regime of high energy QCD, while in the present case for holographic quark-gluon plasma, which has been thought of as strongly-coupled one, we resort to its gravity dual for calculations of above quantities. There is a standard prescription \cite{0205051} for calculating the retarded Green function such as Eq.~(\ref{polarization2}) under the approach of gauge/gravity duality. The current $J_{\mu}$ couples to its source $A_{\mu}\left(x,r=0\right)$ as
\begin{equation}
S_{int}=\int d^4x J_{\mu} A^{\mu}\left(x,r=0\right),
\end{equation}
where $A_{\mu}\left(x,r\right)$ is the flavor $U(1)$ gauge field introduced in next subsection. The main idea under this prescription is to invert the operator Green function on field theory side to dual field Green function on gravity side, whose calculations just need classical gravity action. More details on this prescription can be found in \cite{0205051}. Now we explicitly write down the expression for the polarization tensor defined in Eq.~(\ref{polarization2}) in terms of the variables on dual gravity side
\begin{equation}
\Pi_{\mu\nu}\left(k,T\right)\equiv \frac{\delta^2 S}{\delta A_
{\mu}(-k)\delta A_{\nu}(k)} \mid_{r=0},
\end{equation}
where $S$ is the on-shell action defined in Eqs.~(\ref{D8 action}) and (\ref{D4 action}). Later, it will be found that due to coupling of $A_x$ and $A_t$, we have to express the on-shell action in terms of longitudinal mode $A_L$ and transverse modes $A_y, A_z$ defined in next subsection.

\subsection{Basic equations: flavor current propagation in plasma}
As in many works on applications of gauge/gravity duality to strong coupling problems, we will choose radial gauge for gauge potential, i.e., $A_r=0$. This is enough for non-critical case while for the Sakai-Sugimoto model we also have internal symmetry on $\Omega_4$ space. For brevity, we also set gauge potential along it to zero and assume that gauge potential does not depend on the internal coordinate. As to the action, we just retain to quadratic order in the gauge field fluctuation, which is enough for the propagation of the flavor current. In the following, we will just write down the main equations for the gauge fields (for details on the extraction of them, see \cite{PhysRevD.77.126008,1105.3646}).

For the Sakai-Sugimoto model, the DBI action for the fluctuation of the flavor U(1) gauge field takes the following form after integrating out the $\Omega_4$ space:
\begin{eqnarray}
&&
S_8=-\frac{\left(2\pi l_s^2\right)^2}{4}T_8 N_f V_{S^4}\int d^4xdr
 \sqrt{-g_{eff}}g^{MN}g^{PQ}F_{MP}F_{NQ},\\
&&
\sqrt{-g_{eff}}\equiv e^{-\phi}\sqrt{-\det g_5}g^2_{S^4} = g_s^{-1}R^{\frac{3}{2}}u_T^{\frac{7}{2}}r^{-\frac{9}{2}}
\end{eqnarray}
where $F_{MP}=\partial_MA_P-\partial_PA_M$ is the field strength of the gauge field $A_M$, $g^{MN}$ is the inversion of the induced metric in Eq.~(\ref{induced metric for ss}), the index $M,N$ and so on just need to run the former five ones in Eq.~(\ref{induced metric for ss}), $T_8$ is the D8 brane tension and $V_{S^4}$ is the volume of unit sphere $\Omega_4$. The equation of motion (EOM) for the gauge field $A_M$ followed from this action is of Maxwell-type,
\begin{equation}
\partial_M\left(\sqrt{-g_{eff}}g^{MN}g^{PQ}F_{NQ}\right)=0.
\label{eom}
\end{equation}

Now we turn to the momentum space by doing the following partial Fourier transformation of the gauge field
\begin{equation}
A_\mu\left(x,r\right)=\displaystyle\int\frac{d^4k}{(2\pi)^4}e^{ikx}A_\mu
\left(k_\mu,r\right),\quad k_\mu=\left(-\omega,q,0,0\right).\label{eq:A}
\end{equation}
Without loss of generality, we have chosen the spatial momentum along just one spatial direction as done in the literature. Then in the partial momentum space, the on-shell action turns into the following form,
\begin{eqnarray}
S_8&=&-\displaystyle\frac{\left(2\pi l_s^2\right)^2}{2}T_8N_fV_{S^4}\displaystyle\int
\frac{d^4k}{\left(2\pi\right)^4 }\sqrt{-g_{eff}}g^{rr}\left\{g^{tt}A_t
\left(-k,r\right)\partial_rA_t\left(k,r\right)
\right.\nonumber\\
&& \left.~+g^{ii}A_i\left(-k,r\right)\partial_rA_i\left(k,r\right)\right\}\mid^{r=1}_{r=0} \qquad
\left(i=x,y,z\right),\label{D8 action}
\end{eqnarray}
while Eq.~(\ref{eom}) can be explicitly casted into following three ones
\begin{eqnarray}
&& \omega A^\prime_t+q f(r) A^\prime_x=0,\label{eom of at ax} \\
&&
A^{\prime\prime}_y+\frac{\partial_r\left(\sqrt{-g_{eff}}g^{rr}g^{yy}\right)}
{\sqrt{-g_{eff}}g^{rr}g^{yy}}A^{\prime}_y-\frac{g^{yy}}{g^{rr}}
\left(q^2-\frac{\omega^2}{f(r)}\right)A_y=0,\label{eom of ay} \\
&& A_t^{\prime\prime\prime}+\left[\frac{2\partial_r\left(\sqrt{-g_{eff}}g^{rr}g^{tt}
\right)}{\sqrt{-g_{eff}}g^{rr}g^{tt}}-\frac{\partial_r\left(\sqrt{-g_{eff}}g^{tt}g^{xx}
\right)}{\sqrt{-g_{eff}}g^{tt}g^{xx}}\right]A_t^{\prime\prime}+\left\{\frac{\partial^2_r
\left(\sqrt{-g_{eff}}g^{rr}g^{tt}\right)}{\sqrt{-g_{eff}}g^{rr}g^{tt}}\right.\nonumber\\
&&~~~~\left.~-\frac{g^{xx}}{g^{rr}}\left(q^2-\frac{\omega^2}{f(r)}\right)-
\frac{\partial_r\left(\sqrt{-g_{eff}}g^{tt}g^{xx}\right)}{\sqrt{-g_{eff}}g^{tt}g^{xx}}
\frac{\partial_r\left(\sqrt{-g_{eff}}g^{rr}g^{tt}\right)}{\sqrt{-g_{eff}}g^{rr}g^{tt}}
\right\}A_t^{\prime}=0.\label{eom of at}
\end{eqnarray}
The component $A_z$ satisfies the same equation as the $A_y$-component in Eq.~(\ref{eom of ay}), and we refer to them as transverse modes. Another important relation between $A_x$ and $A_t$ is
\begin{equation}
\partial_r\left(\sqrt{-g_{eff}}g^{rr}g^{tt}A_t^{\prime}\right)-\sqrt{-g_{eff}}g^{tt}
g^{xx}q\left(\omega A_x+q A_t\right)=0. \label{al}
\end{equation}
In above equations, prime denotes derivative with respect to the radial coordinate $r$.
Once the induced metric was inserted into Eqs.~(\ref{eom of ay},\ref{eom of at},\ref{al}),
they can be simplified to the following forms
\begin{eqnarray}
&& a^{\prime\prime}+\left[\frac{1}{2r}+\frac{f^{\prime}(r)}{f(r)}\right]a^{\prime}+\left[
\frac{\tilde{\omega}^2-\tilde{q}^2f(r)}{rf^2(r)}-\frac{f^{\prime}(r)}{2rf(r)}\right]a=0,
\label{eom of a}\\
&&
A_y^{\prime\prime}+\left[-\frac{1}{2r}+\frac{f^{\prime}(r)}{f(r)}\right]A_y^{\prime}-
\frac{\tilde{q}^2f(r)-\tilde{\omega}^2}{rf^2(r)}A_y=0,\label{eom of ay1} \\
&&
A_L=\frac{u_T}{R^3}q^{-1}r^{\frac{3}{2}}f(r)\left(r^{-1/2}a\right)^{\prime}\label{al1},
\end{eqnarray}
where $(\tilde{\omega},\tilde{q})\equiv\frac{3}{4\pi T}(\omega,q)$ are dimensionless variables, $a$ denotes $A_t^{\prime}$, and $A_L$ is the longitudinal mode defined as $A_L\equiv q A_t+\omega A_x$. Note that we have written Eqs.~(\ref{eom of at ax},\ref{eom of ay},\ref{eom of at},\ref{al}) in a general form so that they should also be suitable for the non-critical model being studied later.

Similarly, we can write down similar equations for the non-critical version of the Sakai-Sugimoto model, but for brevity we just list the main results like Eqs.~(\ref{D8 action},\ref{eom of a},\ref{eom of ay1},\ref{al1}):
\begin{eqnarray}
&& S_4=-\displaystyle\frac{\left(2\pi l_s^2\right)^2}{2}
\frac{T_4 N_f}{e^{\phi}}\frac{u_T^2}{R^3}\displaystyle\int \frac{d^4k}{\left(2\pi\right)^4}
\frac{1}{r}\left[f(r)A_i\left(-k,r\right)\partial_rA_i
\left(k,r\right)\right.\nonumber\\
&& \left. ~~~~~~
-A_t\left(-k,r\right)\partial_rA_t\left(k,r\right)\right]|_{r=0}^{r=1}\qquad \left(i=x,y,z\right),
\label{D4 action}\\
&&
a^{\prime\prime}+\left[\frac{f^{\prime}(r)}{f(r)}-\frac{1}{r}\right]a^{\prime}+
\left[\frac{1}{r^2}-\frac{f^{\prime}(r)}{r f(r)}+\frac{\tilde{\omega}^2
-\tilde{q}^2f(r)}{f^2(r)}\right]a=0,\\
&&
A_y^{\prime\prime}+\left[\frac{f^{\prime}(r)}{f(r)}-\frac{1}{r}\right]A_y^{\prime}+
\frac{\tilde{\omega}^2-\tilde{q}^2f(r)}{f^2(r)}A_y=0,\\
&&
A_L\equiv\frac{u_T^2}{R^4}\frac{1}{q}r f(r)\left(\frac{a}{r}\right)^{\prime}.
\end{eqnarray}
Here the only difference from the Sakai-Sugimoto model in notations lies in the specific definitions of the dimensionless variables $\tilde{\omega}$ and $\tilde{q}$:
\begin{equation}
\tilde{\omega}\equiv\frac{\omega}{0.8\pi T},\qquad \tilde{q}\equiv\frac{q}{0.8\pi T}.
\end{equation}
We now express the flavor brane actions in terms of $A_L$,$A_y$, $A_z$ and $a$ as mentioned in last subsection. Because similar calculations have been done in the literature for many times, we here merely list the final results. For the Sakai-Sugimoto model we have
\begin{eqnarray}
&S_8=-\displaystyle\frac{\left(2\pi l_s^2\right)^2}{2} T_8 N_f V_{S^4}
\displaystyle\int \frac{d^4k} {\left(2\pi\right)^4} \sqrt{-g_{eff}} g^{rr}
\left\{g^{tt} q^{-1}A_L\left(-k,r \right)a\left(k,r\right)\right.\nonumber\\
&\left. +g^{ii} A_i\left(-k,r\right)\partial_r A_i\left(k,r\right)\right\}\mid_{r=0}^{r=1}
\qquad \left(i=y,z\right),\label{D8 on shell}
\end{eqnarray}
and for the non-critical model
\begin{eqnarray}
&S_4=-\displaystyle\frac{\left(2\pi l_s^2\right)^2}{2} \frac{T_4 N_f}{e^{\phi}}
\frac{u_T^2}{R^3} \displaystyle\int \frac{d^4k} {\left(2\pi\right)^4} r^{-1}
\left\{- q^{-1}A_L\left(-k,r \right)a\left(k,r\right) \right.\nonumber\\
&\left. +f(r) A_i\left(-k,r\right)\partial_r A_i\left(k,r\right)\right\}\mid_{r=0}^{r=1} \qquad
\left(i=y,z\right).\label{D4 on shell}
\end{eqnarray}
Some remarks are due about these equations which will determine later calculations.
Explicitly, these equations are more involved than the related ones in Ref.~\cite{0710.5297}, which has already been confirmed in \cite{0912.0231,*0912.2238} for D3/D7 brane setup. We can attribute these complicatedness to the introduction of the flavor branes and the non-conformality of the model. However, once these EOMs were turned into standard Schr$\ddot{\text{o}}$dinger types and taking high momentum and high frequency limits, we will find that these complications will automatically disappear and they are qualitatively in common with those of \cite{0710.5297}.

We now proceed by following \cite{0710.5297} to turn these EOMs into standard
Schr$\ddot{\text{o}}$dinger types and discuss some general features of effective
potentials respectively. We will find that these general discussions for potentials will
reveal some physical pictures for the DIS processes considered in this note.

The equations obeyed by new fields have the following common form irrespective of the models
\begin{eqnarray}
&&
-\psi_L^{\prime\prime}+V_L(r)\psi_L=0 \qquad \text{(for time component $a$)},\label{schl}\\
&&
-\psi_T^{\prime\prime}+V_T(r)\psi_T=0 \qquad \text{(for transverse component $A_y$ or $A_z$)}.
\label{scht}
\end{eqnarray}
For the Sakai-Sugimoto model, the explicit field transformations and effective potentials are as follows
\begin{eqnarray}
&&
a=\sqrt{\frac{1}{\left(1-r^3\right)r^{\frac{1}{2}}}}\psi_L,\label{a}\\
&&
V_L(r)=\frac{1}{r\left(1-r^3\right)^2}\left[\frac{1}{16r}\left(-3-78r^3+45r^6\right)
+ K^2-\tilde{q}^2 r^3\right];\\
&&
A_y=\sqrt{\frac{r^{\frac{1}{2}}}{1-r^3}}\psi_T,\label{ay}\\
&&
V_T(r)=\frac{1}{r\left(1-r^3\right)^2}\left[\frac{1}{16r}\left(5-46r^3+5r^6\right)+K^2
-\tilde{q}^2 r^3\right].
\end{eqnarray}
While for the non-critical model, we have following respective ones
\begin{eqnarray}
&& a=\sqrt{\frac{r}{\left(1-r^5\right)}}\psi_L,\\
&&
V_L(r)=\frac{1}{\left(1-r^5\right)^2}\left[\frac{1}{4r^2}\left(-1-48r^5+24r^{10}\right)
+K^2-\tilde{q}^2 r^5\right];\\
&&
A_y=\sqrt{\frac{r}{\left(1-r^5\right)}}\psi_T,\\
&&
V_T(r)=\frac{1}{\left(1-r^5\right)^2}\left[\frac{1}{4r^2}\left(3+8r^{10}-36r^5\right)
+K^2-\tilde{q}^2 r^5\right].
\end{eqnarray}
In these equations, we have defined the dimensionless current virtuality as $K^2\equiv\tilde{q}^2-\tilde{\omega}^2$. We will take this virtuality to be space-like, which amounts to saying that the process considered here is like the lepton deep inelastic scattering off the proton. But as noted in \cite{0710.5297}, the final results will also be suitable for time-like virtuality if we take high momentum limit (especially if $\tilde{q}^2\gg K^2$). Since we are interested in the internal structure of holographic quark-gluon plasma, we should use high energy probe to explore this just as the DIS processes in pQCD. One basic difference between the plasma and a single hadron is that the former has an intrinsic scale (temperature). In short, we are focusing on the following kinematic
parameter space:
\begin{equation}
\tilde{\omega}\gg 1,\qquad \tilde{q}\gg 1,\qquad K^2\gg 1.\label{kinematics}
\end{equation}
Under the above kinematics, these effective potentials can be further approximated as follows\footnote{For the following four potentials, we have ignored polynomials of $r$ in their final expressions. Strictly speaking, this is not right; but since we are in particular concerning their behavior near the boundary $r=0$, which directly determines the polarization tensor, and their behavior at the horizon $r=1$ is just for incoming wave boundary condition, we believe this is reasonable.}. For the Sakai-Sugimoto model we have
\begin{eqnarray}
&&
V_L(r)=\frac{1}{r\left(1-r^3\right)^2}\left[-\frac{3}{16r}+K^2
-\tilde{q}^2 r^3\right],\label{vl for ss}\\
&&
V_T(r)=\frac{1}{r\left(1-r^3\right)^2}\left[\frac{5}{16r}+K^2
-\tilde{q}^2 r^3\right]; \label{vt for ss}
\end{eqnarray}
while for the non-critical one
\begin{eqnarray}
&&
V_L(r)=\frac{1}{\left(1-r^5\right)^2}\left[-\frac{1}{4r^2}+K^2-\tilde{q}^2 r^5\right],\label{vl for non-ss}\\
&&
V_T(r)=\frac{1}{\left(1-r^5\right)^2}\left[\frac{3}{4r^2}+K^2-\tilde{q}^2 r^5\right]. \label{vt for non-ss}
\end{eqnarray}
Eqs.~(\ref{schl},\ref{scht},\ref{vl for ss},\ref{vt for ss},\ref{vl for non-ss},\ref{vt for non-ss}) are the main ingredients for later extraction of the structure functions of holographic plasma.

We have seen that, in the interesting kinematic regime symbolled by Eq.~(\ref{kinematics}), for both the Sakai-Sugimoto model and its non-critical version, the effective potentials\footnote{We should have plotted these effective potentials for illustration as in \cite{0710.5297,0912.0231,*0912.2238}, but we skip it here for brevity since they are basically the same.} for longitudinal as well as transverse modes are qualitatively similar to those of the $\mathcal{N}=4$ SYM case with(out) flavors \cite{0710.5297,0912.0231,*0912.2238}. More explicitly, the maximum of longitudinal potential can be positive (corresponding to potential barrier), negative (with no barrier at all) or zero according to the value of $\tilde{q}/K^4$ (Sakai-Sugimoto model) or $\tilde{q}^2/K^7$ (non-critical model). While effective potentials for the transverse modes are more involved because they start from a positive infinity and then falling to negative infinity very rapidly, which may complicate later analysis using WKB approximation to construct the wave function $\psi_T$. Recalling the fact that the dilaton vacuum for the Sakai-Sugimoto model is not a constant while for the non-critical version it is constant, these facts together seem to indicate that non-conformality of the Sakai-Sugimoto model is not essential on physical picture of high energy scattering process. This is a byproduct of the general behavior analysis for the effective potentials, later we will confirm this observation by direct extraction of the structure functions for both holographic models.

Before concluding this section, we briefly summarize the physical picture in \cite{0710.5297} governing high energy DIS from the viewpoint of non-relativistic quantum mechanics. For longitudinal mode, when the potential barrier builds in (corresponding to small spatial momentum case), the wave function cannot be imposed incoming wave condition at the horizon due to the high and narrow barrier, which indicating zero structure function. When taking into account non-perturbative tunneling effect, a small exponentially suppressed structure function can be obtained. We in following sections therefore merely focus on high spatial momentum limit. In this regime, the wave function
will be complex one and incoming wave condition can be imposed at the horizon. Moreover, in this high energy regime, a partonic picture for the plasma exists.

\section{Structure function of holographic quark-gluon plasma}\label{section3}
Now we have all the elements to calculate the polarization tensor defined in Eq.~(\ref{polarization2}). As mentioned above, we should focus on the high momentum kinematics besides the one defined in Eq.~(\ref{kinematics}). This means that the $K^2$-terms in effective potentials can be ignored as well, which makes semi-analytical solutions for these Schr$\ddot{\text{o}}$dinger equations possible. We in this section follow the standard WKB approach in non-relativistic quantum mechanics to construct these solutions. We present the calculations in detail for the Sakai-Sugimoto model and then list the final results for its non-critical version.

We first discuss the longitudinal mode. Under high momentum approximation discussed in last paragraph, Schr$\ddot{\text{o}}$dinger equation for longitudinal mode then takes following simple form near $r\simeq0$:
\begin{equation}
\psi_L^{\prime\prime}(r)+\left(\frac{3}{16r^2}+\tilde{q}^2 r^2\right)\psi_L(r)=0.
\end{equation}
Its general solution is a linear combination of the Bessel and Neumann functions:
\begin{equation}
\psi_L\left(r\simeq 0\right)=c_1 \tilde{q}^{1/4}r^{1/2} J_{1/8}
\left(\frac{\tilde{q}r^2}{2}\right) +c_2 \tilde{q}^{1/4}r^{1/2} N_{1/8}\left(\frac{\tilde{q}r^2}{2}\right),
\end{equation}
and the constants $c_1,c_2$ will be determined by imposing incoming wave boundary condition at the horizon, which needs do matches of solutions at different regimes.

Near $r\simeq1$, Schr$\ddot{\text{o}}$dinger equation can be approximated as
\begin{equation}
\psi_L^{\prime\prime}(r)+\frac{\tilde{q}^2}{9\left(1-r\right)^2}\psi_L(r)=0.
\end{equation}
Its general solution is
\begin{equation}
\psi_L\left(r\simeq1\right)=c_3 \left(1-r\right)^{\frac{1}{2}-i\frac{\tilde{q}}{3}}
+c_4 \left(1-r\right)^{\frac{1}{2}+i\frac{\tilde{q}}{3}}.
\end{equation}
Recalling Eqs.~(\ref{a},\ref{ay}) and imposing incoming wave boundary condition at the horizon make us to conclude that $c_4=0$, thus leaving the general solution near the horizon to be
\begin{equation}
\psi_L\left(r\simeq1\right)=c_3 \left(1-r\right)^{\frac{1}{2}-i\frac{\tilde{q}}{3}}.
\end{equation}

Besides, we have to study in detail the solution suitable for intermediate regime far
from the singularities at $r=0$ and $r=1$. In this regime, Schr$\ddot{\text{o}}$dinger equation is approximated as
\begin{equation}
\psi_L^{\prime\prime}(r)+\frac{\tilde{q}^2 r^2}{\left(1-r^3\right)^2}\psi_L(r)=0.
\end{equation}
For convenience, we now define the so-called canonical momentum $p(r)$ and action $s(r)$:
\[p(r)=\frac{\tilde{q}r}{1-r^3},\qquad s(r)=\int^r_0 p(r)dr=\int^r_0 dr\frac{\tilde{q}r}{1-r^3}.\]
Then the two linear independent solutions in intermediate regime under WKB approximation are
\begin{eqnarray}
&&
\psi_L^{(1)}(r)=\frac{1}{\sqrt{p(r)}}\cos\left[s(r)+\phi_1\right],\\
&&
\psi_L^{(2)}(r)=\frac{1}{\sqrt{p(r)}}\sin\left[s(r)+\phi_2\right].
\end{eqnarray}
The asymptotic behaviors for $p(r)$ and $s(r)$ at singularities are necessary for the solution matching underlying different regimes
\begin{eqnarray}
&&
p\left(r\simeq0\right)\simeq \tilde{q}r \qquad \text{and}\qquad
p\left(r\simeq1\right)\simeq \frac{\tilde{q}}{3(1-r)},\\
&&
s\left(r\simeq0\right)\simeq \frac{\tilde{q}r^2}{2} \qquad \text{and} \qquad
s\left(r\simeq1\right)\simeq -\frac{\tilde{q}}{3}\log\left(1-r\right)+\text{constant}.
\end{eqnarray}

The next step is to match these solutions to determine $c_1$, $c_2$ and $c_3$. In doing this, we need the asymptotic expansion for the Bessel or Neumann function for very large variable\footnote{Although $r$ is small here, $\tilde{q}$ is large and we therefore regard $\tilde{q}r^2/2$ to be large enough.}
\begin{eqnarray}
&&
\displaystyle J_{1/8}\left(\frac{\tilde{q}r^2}{2}\right)\simeq
\sqrt{\frac{4}{\pi \tilde{q}r^2}} \cos\left(\frac{\tilde{q}r^2}{2}-\frac{\pi}{16}-\frac{\pi}{4}\right),\\
&&
\displaystyle N_{1/8}\left(\frac{\tilde{q}r^2}{2}\right)\simeq \sqrt{\frac{4}{\pi \tilde{q}r^2}}
\sin\left(\frac{\tilde{q}r^2}{2}-\frac{\pi}{16}-\frac{\pi}{4}\right).
\end{eqnarray}
It is then easily found that if we choose $\phi_1=\phi_2=-5\pi/16$ and $c_2=i c_1$,
the matching of solutions in different regimes accomplishes. Note that the condition $c_2=i c_1$ is the main result from the solution matching, which is also a direct reflection of the incoming wave boundary condition imposed at the horizon.

Now we easily arrive at the boundary behavior for the wave function $\psi_L(r)$ as
\begin{eqnarray}
\psi_L\left(r\simeq 0\right)&=&c_1 \tilde{q}^{1/4}r^{1/2} J_{1/8}
\left(\frac{\tilde{q}r^2}{2}\right) +i c_1 \tilde{q}^{1/4}r^{1/2} N_{1/8}
\left(\frac{\tilde{q}r^2}{2}\right)\nonumber\\
&\equiv&c_1 \tilde{q}^{1/4}r^{1/2} H^{(1)}_{1/8}\left(\frac{\tilde{q}r^2}{2}\right).
\end{eqnarray}
In the second line of above equation, we have written the solution as the first kind
Hankel function with order $1/8$.

The effective potential for transverse mode is more involved, but the analysis under WKB approximation is similar, therefore we here just list the final results. Matching of solutions in three different regimes (near horizon, near boundary and intermediate regime far from these singularities) results in the following boundary behavior for transverse mode $A_y$ and $A_z$
\begin{equation}
\psi_T\left(r\simeq0\right)=c_1^i \tilde{q}^{1/4}\sqrt{\frac{r^{3/2}}{1-r^3}}H_{3/8}^{(1)}\left(\tilde{q}r^2/2\right).
\end{equation}

The only undetermined constants $c_1$ and $c_1^i$ can be expressed in terms of boundary values of the gauge field $A_L(r=0)\equiv A_L(0)$ or $A_{y,z}(r=0)\equiv A_i(0)$ respectively. This can be achieved by using Eqs.~(\ref{al1},\ref{a},\ref{ay}) and the final results essential for later extraction of structure functions are listed as below.
For longitudinal mode we have
\begin{eqnarray}
&&
A_L(r\simeq0)=c_1 \frac{u_T}{R^3}q^{-1} \tilde{q}^{1/4}r^{3/2}(1-r^3) \left\{\sqrt{\frac{1}{(1-r^3)r^{1/2}}}H^{(1)}_{1/8}\left(
\frac{\tilde{q}r^2}{2}\right)\right\}^{\prime},\\
&&
a(r\simeq0)=c_1 \tilde{q}^{1/4} \sqrt{\frac{r^{1/2}}{(1-r^3)}}H^{(1)}_{1/8}\left(\frac{\tilde{q}r^2}{2}\right),\\
&&
c_1=\frac{-i\pi 2^{3/4} q} {\Gamma(1/8)\tilde{q}^{1/8}}\frac{R^3}{u_T}A_L(0);
\end{eqnarray}
and for transverse modes $A_i$ $(i=y,z)$ we have
\begin{eqnarray}
&&
A_i(r\simeq0)=c_1^{i}\tilde{q}^{1/4}\sqrt{\frac{r^{3/2}}{1-r^3}}H^{(1)}_{3/8}
\left(\frac{\tilde{q}r^2}{2}\right),\\
&&
c_1^{i}=\frac{i\pi \tilde{q}^{1/8}}{2^{3/4}\Gamma(3/8)}A_i(0).
\end{eqnarray}
With these solutions, we can derive the expressions for the thermal polarization tensor
defined in Eq.~(\ref{polarization2}):
\begin{eqnarray}
&&
\text{Im}\Pi_{LL}\left(k,T\right)%=\frac{\left(2\pi l_s^2\right)^2T_8 N_f V_{S^4}R^6}{2
%g_s}\frac{64\sqrt{2}\pi^2}{3\Gamma^2(1/8)}T\tilde{q}^{1/4}
=\frac{\sqrt{2}l_s}{12\Gamma^2(1/8)g_s}\lambda N_f N_c T \tilde{q}^{1/4},\\
&&
\text{Im}\Pi_{yy}\left(k,T\right)=\text{Im}\Pi_{zz}\left(k,T\right)%=\frac{\left(2\pi
%l_s^2 \right)^2T_8 N_f V_{S^4}R^6}{2 g_s}
%\frac{128\sqrt{2}\pi^5}{27\Gamma^2(1/8)}T^3\tilde{q}^{3/4}
=\frac{\sqrt{2}\pi l_s}{54\Gamma^2(3/8)g_s}\lambda N_f N_c T^3 \tilde{q}^{3/4},
\end{eqnarray}
and other components are exactly vanishing. Referring to above results for the polarization tensor, we have just listed the imaginary parts which are directly related to the structure function. In fact, the real part of polarization tensor are divergent and therefore needs regularized (here we skipped these details). Moreover, we here for simplification assume the regularization will not introduce new terms into the imaginary parts. Therefore, the structure functions $F_1$ and $F_2$ are easily derived via Eqs.~(\ref{F1},\ref{F2},\ref{F2'})
\begin{eqnarray}
F_1\left(k,T\right)&=&\frac{\sqrt{2}l_s}{108\Gamma^2(3/8)g_s}\lambda N_f N_c T^3
\tilde{q}^{3/4}\simeq\frac{l_s}{g_s}\lambda N_f N_c T^3 \tilde{q}^{3/4},\label{F1 for ss}\\
F_2\left(k,T\right)&=&\frac{\omega^2}{q^2}\left[\frac{Q^2 x}{\pi}\frac{\sqrt{2}l_s}
{12\Gamma^2(1/8)g_s}\lambda N_f N_c T \tilde{q}^{1/4}+2x F_1\right]\nonumber\\
&\simeq&%\frac{l_s}{g_s}\lambda N_f N_c \left(Q^2 x T\tilde{q}^{1/4}+x T^3
%\tilde{q}^{3/4}\right)
2xF_1.\label{F2 for ss}\label{F2 for ss}
\end{eqnarray}
In the first line of Eq.~(\ref{F2 for ss}) one can easily show that the first term can be ignored when comparing to the second one in the interesting kinematic regime $\tilde{q}/K^4\gg1$. Following \cite{0710.5297}, we now express the two structure functions in terms of the Bjoken variable $x$ defined in Sec.~\ref{sub DIS} and the flavor current virtuality $Q^2$ as
\begin{eqnarray}
&&F_1(x,Q^2)=\frac{\sqrt{2}l_s}{108\Gamma^2(3/8)g_s}\lambda N_f N_c T^3
\left(\frac{3Q^2}{8\pi x T^2}\right)^{3/4}\sim\lambda N_f N_c T^3
\left(\frac{3Q^2}{8\pi x T^2}\right)^{3/4},\label{F1' for ss}\\
&&
F_2(x,Q^2)\simeq 2x F_1(x,Q^2)\sim 2\lambda N_f N_c T^3 x
\left(\frac{3Q^2}{8\pi x T^2}\right)^{3/4}.\label{F2' for ss}
\end{eqnarray}
In obtaining the above two equations, we have used one approximate relation $\tilde{\omega}\simeq\tilde{q}$ to express the spatial momentum as $q\simeq Q^2/(2x T)$.

Before closing this subsection, we briefly carry out similar analysis for the non-critical model. Since in the previous subsection the general procedure for using WKB method to solve Schr$\ddot{\text{o}}$dinger problem has been presented in detail, we here just list corresponding key results. The solution for longitudinal mode near boundary $r=0$ behaves as
\begin{eqnarray}
&&
A_L(r\simeq0)=C_1\frac{u_T^2}{R^4}q^{-1}\tilde{q}^{1/7}r(1-r^5)
\left\{\sqrt{\frac{1} {1-r^5}}H_{0}^{(1)} \left(\frac{2}{7}\tilde{q}r^{7/2}\right)\right\}^{\prime},\\
&&
a(r\simeq0)=C_1 \tilde{q}^{1/7}\frac{r}{\sqrt{{1-r^5}}} H_{0}^{(1)}\left(\frac{2}{7}\tilde{q}r^{7/2}\right),\\
&&
C_1=-\frac{i\pi}{7}\frac{q}{\tilde{q}^{1/7}}\frac{R^4}{u_T^2}A_L(0);
\end{eqnarray}
and for transverse mode $A_i(r)$ $(i=y,z)$
\begin{eqnarray}
&&
A_i(r\simeq0)=C_1^{i}\tilde{q}^{1/7}\frac{r}{\sqrt{1-r^5}}H_{2/7}^{(1)}\left(\frac{2}{7} \tilde{q}r^{7/2}\right),\\
&&
C_1^i=\frac{i\pi\tilde{q}^{1/7}}{7^{2/7}\Gamma(2/7)}A_i(0).
\end{eqnarray}

The polarization tensor can be derived by inserting these solutions into the on-shell action Eq.~(\ref{D4 on shell})
\begin{eqnarray}
&&
\text{Im}\Pi_{LL}(k,T)%=\frac{\left(2\pi l_s^2\right)^2 T_4 N_f\pi R}{7 e^{\phi}}
=\frac{\sqrt{3}R N_f N_c}{56\sqrt{2}\pi g_s l_s},\\
&&
\text{Im}\Pi_{yy}(k,T)%=\text{Im}\Pi_{zz}(k,T)=\frac{\left(2\pi l_s^2\right)^2 T_4 N_f}{
%e^{\phi}}\frac{142\pi^3 R}{25\Gamma^2(2/7)}T^2 \left(\frac{\tilde{q}}{7}\right)^{4/7}
=\frac{7\sqrt{6}\pi R N_f N_c}{25\Gamma^2(2/7)g_s l_s}T^2 \left(\frac{\tilde{q}}{7}\right)^{4/7}.
\end{eqnarray}
Then the structure functions are
\begin{eqnarray}
F_1\left(k,T\right)&=&\frac{7\sqrt{6} R N_f N_c}{50\Gamma^2(2/7)g_s l_s}T^2
\left(\frac{\tilde{q}}{7}\right)^{4/7}\simeq N_f N_c T^2
\left(\frac{\tilde{q}}{7}\right)^{4/7},\label{F1 for non-ss}\\
F_2\left(k,T\right)&=&\frac{\omega^2}{q^2}\left[\frac{Q^2 x}{\pi}
\frac{\sqrt{3}N_f N_c }{56\sqrt{2}\pi g_s l_s}+\frac{7\sqrt{6} R N_f N_c}{25\Gamma^2(2/7)g_s l_s}x T^2
\left(\frac{\tilde{q}}{7}\right)^{4/7}\right]\nonumber\\
&\simeq&2x F_1(k,T).\label{F2 for non-ss}
\end{eqnarray}
In the first line of Eq.~(\ref{F2 for non-ss}), one can easily show that the first term can be neglected in the interesting regime $\tilde{q}^{4/7}\gg K^2$, so we have the approximate equality of the second line as in the Sakai-Sugimoto model. We now express these results in terms of the Bjoken variable $x$ and the flavor current virtuality $Q^2$ as in the Sakai-Sugimoto model
\begin{eqnarray}
&&
F_1(x,Q^2)\simeq \frac{7\sqrt{6}R}{50\Gamma^2(2/7)g_s l_s}N_f N_c T^2
\left(\frac{5Q^2}{8\pi xT^2}\right)^{4/7}\sim N_f N_c T^2
\left(\frac{5Q^2}{8\pi xT^2}\right)^{4/7},\label{F1' for non-ss}\\
&&
F_2(x,Q^2)\simeq 2x F_1(x,Q^2)\sim2N_f N_c T^2 x\left(\frac{5Q^2}{8\pi xT^2}\right)^{4/7}.\label{F2' for non-ss}
\end{eqnarray}

Eqs.~(\ref{F1 for ss},\ref{F2 for ss},\ref{F1' for ss},\ref{F2' for ss},\ref{F1 for non-ss},\ref{F2 for non-ss},\ref{F1' for non-ss},\ref{F2' for non-ss}) comprise the main results of this note. Now we have a short remark on these results and also a brief comparison between these two models as well as the well-investigated $\mathcal{N}=4$ SYM case.

The first point is that we have an analogy of Callan-Gross relation $F_2\simeq2x F_1$ in the interesting kinematic regime considered in this note. In pQCD, this relation holds only at relatively large Bjoken variable $x$, where parton structures of hadrons are dominated by the valence quarks. This relation has been already obtained in DIS off $\mathcal{N}=4$ SYM plasma with(out) flavors and here it also holds for the Sakai-Sugimoto model as well as its non-critical version. Since the setups of holographic dual of sQGP are very general, and what is more is that the physical picture underlying the $\mathcal{R}$ or the flavor current DIS off the plasma is quite simple and general, these surprising facts seem to indicate that it may be a general relation for holographic quark-gluon plasma.

The second key point is concerned with the non-conformal characteristic of the Sakai-Sugimoto model, which is also the main motivation of the present study. Because in our interesting kinematic regime, the two structure functions are related to each other by the Callan-Gross relation, we then mainly focus on the function $F_1$. It is clear that for two models $F_1$ presents scaling behavior on their dependence on temperature $T$ and dimensionless spatial momentum $\tilde{q}$. Their dependence on dimensionless momentum $\tilde{q}$ are approximately the same, while on temperature are quite different: $\sim T^3$ for the Sakai-Sugimoto model and $\sim T^2$ for the non-critical one. Recalling that the latter scaling behavior $\sim T^2$ has also been valid in \cite{0710.5297} for the $\mathcal{N}=4$ SYM plasma, we may think of the essential effect of non-conformality of the Sakai-Sugimoto model as the $T^3$-scaling behavior of the $F_1$ structure function. However, this guess needs further confirmation or cancelation because the non-conformality of the Sakai-Sugimoto model is not well-controlled. Moreover, the gauge coupling constant of strong interaction has a logarithmical running with evolving energy scale and simple gravity realization of this kind of gauge theory has been established by combining top-down and bottom-up approaches in \cite{0707.1324,*0707.1349}, it may be interesting to resort to this kind of model to probe the effect of gauge coupling running on the internal structure of sQGP.

The last point we would like to stress is about the pre-factors for the structure functions. Similar to \cite{0912.0231,*0912.2238}, $N_f N_c$ counts the number of freedom of the plasma, and we here probe the quark sector. The models considered in this note has left some stringy imprints on the quantities of field theoretical side (here they are the plasma structure function). This can be easily read from
Eqs.~(\ref{F1 for ss},\ref{F2 for ss},\ref{F1 for non-ss},\ref{F2 for non-ss}), which have explicit dependence on string coupling constant $g_s$ and string length $l_s$. Moreover, the behavior concerned to these two stringy parameters seems to be different between the Sakai-Sugimoto model and its non-critical version. These facts seem to be
very strange and even unaccepted because we here focused on the field side quantities and they should not show explicit dependence on the parameters on the gravity or string side. Compared to related results of the $\mathcal{N}=4$ SYM plasma, these undesirable behaviors have not come out there, which seems to say that the D3-brane geometry is a much better gravity dual of some QFT in describing field theoretical physical quantities under gauge/gravity duality technique. However, if we recall that the models we take here are different from realistic QCD theory, then it is acceptable that the results are
counter-intuitive from the viewpoint of field theoretical considerations.

\section{Summary and outlook} \label{section4}
In this note, we have used high temperature version of the Sakai-Sugimoto model, a quite successful gravity dual model of QCD-like theory, to explore the internal structure of holographic quark-gluon plasma. The physical process we analyze here is like the well-investigated DIS in standard QCD, but with the scattered proton replaced by the plasma system and the mediated electromagnetic current simulated by the flavor current as in \cite{0912.0231,*0912.2238}. We have seen that the procedure under gauge/gravity duality technique to study DIS off holographic quark-gluon plasma is quite general and
easily promoted to other gauge/gravity duality setups. The result obtained in this note for the structure function under the Sakai-Sugimoto model is quite different from the well-studied $\mathcal{N}=4$ SYM plasma with(out) flavors. This should be regarded as the effect of the non-conformality of the Sakai-Sugimoto model, which is the most important motivation for our present study. To confirm this, we have also chosen the non-critical version of the Sakai-Sugimoto model for a comparative study. We found that the structure functions for the latter model and the $\mathcal{N}=4$ SYM plasma are much alike. The result of this note seems to contradict the intuition from the fact of asymptotic freedom of pQCD. But we should keep in mind that holographic quark-gluon plasma considered in this note is a strongly-coupled system and we should not expect it to behave exactly as the weakly-interacting regime of realistic QCD. In fact, a more realistic holographic
QCD model taking care of the running of the gauge coupling constant was proposed in \cite{0707.1324,*0707.1349} and we expect to use this model to explore the effect of gauge coupling running on the structure of sQGP.

In realistic QCD, $N_f$ and $N_c$ are of $\mathcal {O}(1)$, while the applicability of gravity dual of $SU(N_c)$ gauge theory usually requires a large $N_c$ limit.
Therefore, we cannot directly compare our results with the data from heavy-ion collision. One prescription to overcome this obstacle is to consider the flavor backreaction to the background geometry and then carry out similar calculations in the $N_f/N_c\sim 1$ limit. Although the hadronic matter produced in heavy-ion collision is of high temperature and high density, we in this note only take into account the high temperature element as in the literature. So in this sense, our present models are not so realistic and need to be promoted to high energy and high density quark-gluon plasma case. Fortunately, under the gauge/gravity duality setup, the matter density also has a gravity realization --- $\mathcal{R}$ charge or flavor charge, which can be realized by rotating color brane along the internal space or as the time component of the flavor gauge field respectively. Then, the analysis of DIS off quark-gluon plasma at high density and high temperature can also be carried out by including this element in the model setup. We leave these problems for future investigations.

\begin{acknowledgments}
The author YYB greatly thanks Zhen-Hua Zhang for useful discussions. This work was supported in part by NSFC (Nos. 10821504,10725526).
\end{acknowledgments}

\end{document}